\documentclass{elsart}
\usepackage[dvips]{graphicx}
\usepackage{xspace}
\newcommand{\cer}{\v{C}erenkov\xspace}
\newcommand{\mevcc}{\ensuremath{\mathrm{MeV}/c^2}\xspace}

\newcommand{\gevc}{\ensuremath{\mathrm{GeV}/c}\xspace}

\begin{document}
\begin{frontmatter}
\title{\cer Particle Identification in FOCUS}
The FOCUS Collaboration
\author[davis]{J.M.~Link},
\author[davis]{M.~Reyes\thanksref{atmex}},
\author[davis]{P.M.~Yager},
\author[cbpf]{J.C.~Anjos},
\author[cbpf]{I.~Bediaga},
\author[cbpf]{C.~G\"obel\thanksref{aturug}},
\author[cbpf]{J.~Magnin},
\author[cbpf]{A.~Massafferri},
\author[cbpf]{J.M.~de~Miranda},
\author[cbpf]{I.M.~Pepe\thanksref{atbahia}},
% ...
\address[davis]{University of California, Davis, CA 95616}
\address[cbpf]{Centro Brasileiro de Pesquisas F\'\i sicas, Rio de Janeiro,
RJ, Brazil}
\address[cine]{CINVESTAV, 07000 M\'exico City, DF, Mexico}
\address[cu]{University of Colorado, Boulder, CO 80309}
\address[fnal]{Fermi National Accelerator Laboratory, Batavia, IL 60510}
\address[fras]{Laboratori  Nazionali di Frascati dell'INFN, Frascati, Italy,
      I-00044}
\address[ui]{University of Illinois, Urbana-Champaign, IL 61801}
\address[iu]{Indiana University, Bloomington, IN 47405}
\address[koru]{Korea University, Seoul, Korea 136-701}
\address[milan]{INFN and University of Milano, Milano, Italy}
\address[anc]{University of North Carolina, Asheville, NC 28804}
\address[pavia]{Dipartimento di Fisica Nucleare e Teorica and INFN,
Pavia, Italy}
\address[pr]{University of Puerto Rico, Mayaguez, PR 00681}
\address[sc]{University of South Carolina, Columbia, SC 29208}
\address[ut]{University of Tennessee, Knoxville, TN 37996}
\address[vandy]{Vanderbilt University, Nashville, TN 37235}
\address[wisc]{University of Wisconsin, Madison, WI 53706}

\thanks[atauton]{Present Address: Instituto de Ciencias Nucleares, Universidad
Nacional Aut\'onoma de M\'exico. CP 04510. M\'exico}
\thanks[atmex]{Present Address: Instituto de F\'\i sica y Matematicas,
Universidad Michoacana de San Nicolas de Hidalgo, Morelia, Mich.,
Mexico 58040} \thanks[aturug]{Present Address: Instituto de F\'\i
sica, Facultad de Ingenier\'\i a, Univ. de la Rep\'ublica, Montevideo,
Uruguay} \thanks[atbahia]{Present Address: Instituto de F\'\i sica,
Universidade Federal da Bahia, Salvador, Brazil}
\thanks[atmilan]{Present Address: INFN sezione di Milano, Milano,
Italy} \thanks[atpub]{Present Address: Instituto de F\'{\i}sica,
Universidad Auton\'oma de Puebla, Puebla, M\'exico}
\thanks[atncar]{Present Address: National Center for Atmospheric
Research, Boulder, CO} \thanks[atvandy]{Present Address: Vanderbilt
University, Nashville, TN 37235} \thanks[atpohang]{Present Address:
Pohang University of Science and Technology, Pohang, Korea 790-784}
\thanks[attaejon]{Present Address: Korea Institute of Standards and
Science, P.O.~Box 102, Yusong-Ku, Taejon 305-600, South Korea}
\thanks[atkyung]{Present Address: Center for High Energy Physics,
Kyungpook National University, 1370 Sankyok-dong, Puk-ku, Taegu,
702-701 Korea} \thanks[atcu]{Present Address: University of Colorado,
Boulder, CO 80309} \thanks[atbrescia]{Present Address: Dipartimento di
Chimica e F\'\i sica per l'Ingegneria e per i Materiali, Universita'
di Brescia and INFN sezione di Pavia} \thanks[atromania]{Present
Address: Nat. Inst. of Phys. and Nucl. Eng., Bucharest, Romania}
\thanks[atlucent]{Present Address: Lucent Technology}
\thanks[atadams]{Present Address: Adams State College, Alamosa, CO
81102} \thanks[atintel]{Present Address: Intel
Corporation,Portland,5200 N.E. Elam Young Parkway Hillsboro,
OR 97124}

\nobreak
\begin{abstract}
We describe the algorithm used to identify
charged tracks in the fixed-target charm-photoproduction experiment
FOCUS\@. We begin by describing the new algorithm
and contrast this approach with that used in our
preceding experiment---E687.  We next illustrate the 
algorithm's performance using physics signals. Finally we briefly describe
some of the methods used to monitor the quantum efficiency and noise
of the \cer cells.
\end{abstract}
\end{frontmatter}
\newpage

\section{Introduction}

FOCUS is a fixed-target experiment concentrating on 
the photoproduction of charm that accumulated data at Fermilab from 1996--1997.
It is a considerably upgraded version of a previous experiment, 
E687 \cite{nim}. In FOCUS, a forward multi-particle spectrometer is used to 
measure the interactions of high energy photons on a segmented BeO
target. We obtained a sample of over 1 million fully reconstructed
charm particles in the three major decay modes: $D^0 \rightarrow K^- \pi^+$,
$K^- \pi^+ \pi^- \pi^+$, and $D^+ \rightarrow K^- \pi^+ \pi^+$ 
(and charge conjugates). We will refer to these as ``golden modes''.  

\begin{figure}[h!]
\begin{center}	
	\includegraphics[width=6.in]{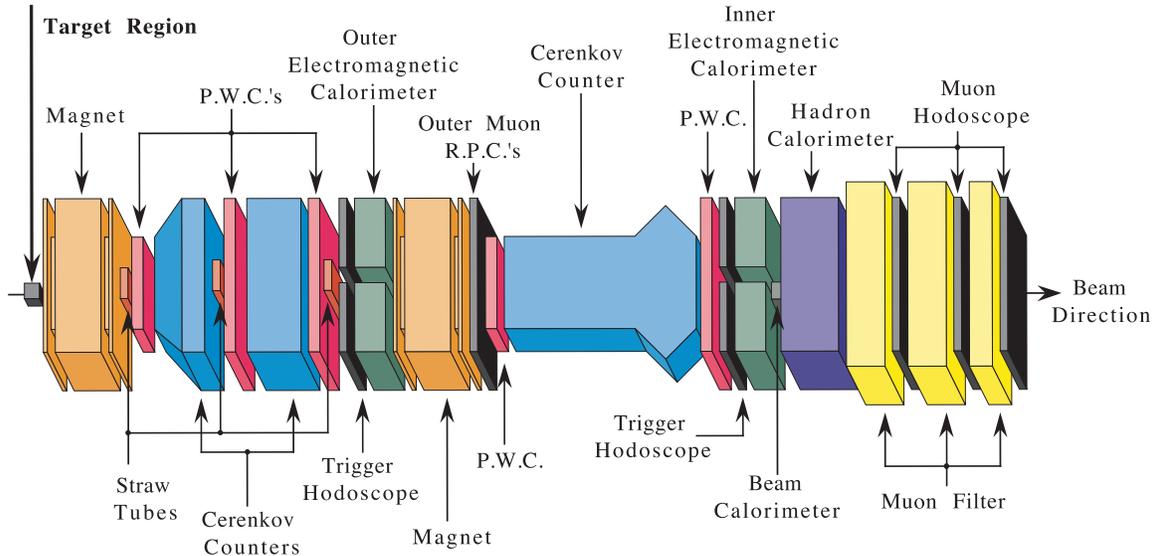}
\end{center} 
\caption{ A schematic drawing of the FOCUS spectrometer. The target
region consists of a segmented BeO target, the interleaved target
silicon planes, trigger counters, and the 12 plane silicon tracking array.
The spectrometer is approximately 32 meters long.
\label{spectrometer} }
\end{figure}

The FOCUS detector (see Figure \ref{spectrometer})
is a large aperture spectrometer with
excellent vertexing and particle identification. A photon beam is
derived from the bremsstrahlung of secondary electrons and positrons
with an $\approx 300$ GeV endpoint energy produced from the 800
GeV/$c$ Tevatron proton beam. The charged particles which emerge from
the target are tracked by two systems of silicon microvertex
detectors. The upstream system, consisting of 4 planes (two views in 2
stations), is interleaved with the experimental target, while the
other system lies downstream of the target and consists of twelve
planes of microstrips arranged in three views. These detectors provide
high resolution separation of primary (production) and secondary
(decay) vertices with an average proper time resolution of $\approx
30~ {\rm fs}$ for 2-track vertices. The momentum of a charged particle
is determined by measuring its deflections in two analysis magnets of
opposite polarity with five stations of multiwire proportional
chambers. 

\section{The FOCUS \cer system}

Three multicell threshold \cer counters are used to
discriminate between electrons, pions, kaons, and protons.  
The \cer system hardware was essentially unchanged from that used
in E687 and is described in detail in Reference \cite{nim}. A brief 
description of the \cer system follows.

There are three multicell thres	hold detectors in the experiment,
referred to as C1, C2, and C3. 	The detectors are operated at	
atmospheric pressure and in the threshold mode.  The gases are chosen
so that different indices of refraction (\emph{i.e.} different light
velocities) establish different momenta in which pions, kaons, and
protons will begin to radiate \cer light (see Table \ref{t:cer}).
For our system the three pion thresholds were chosen to be 4.5, 8.4,
and 17.4~\gevc by use of appropriate gas mixtures. The
photoelectron yield ranged from roughly 2.5 to 20 depending on the
phototube and \cer counter.

The detector C1 is the most upstream of the three \cer counters, lying
just beyond the first analysis magnet, between the first two PWC's
(multiwire proportional chambers) P0 and P1. The gas used was a
helium-nitrogen mixture, and the total length of the gas volume along
the beam direction is 180 centimeters. The \cer detector C2 has the lowest
threshold of the three detectors with a pion threshold of
4.5~\gevc. The gas was pure N$_2$O, and the total length of the
counter gas volume along the beam direction is 188 centimeters. The detector is
located between P1 and P2.

The C1 and C2 \cer counters can detect all charged tracks that are
generally reconstructible in FOCUS.  The 3rd \cer detector C3 is
located downstream of the second analysis magnet. Only higher momentum
tracks make it through the aperature of this magnet, so C3 only helps
in the identification of these tracks. The counter is a helium
threshold counter which was 704 centimeters in length.

\begin{table}[tbp]
\begin{center}
\caption[\cer counter specifications]{\cer counter specifications. The
momentum threshold for the three most relevant charged particles and the
\cer cone radius for a $\beta = 1$ track at the image plane are given for each counter.}
\begin{tabular}{|c||c|c|c|c|c|c|c|}\hline
counter & Gas & \multicolumn{3}{|c|}{Thresh (\gevc)} & No. Cells &  Avg. PE & \cer Radius   \\
        &     & pion  & kaon & proton  &   & & (cm) \\ \hline \hline
C2      & N$_2$O   & 4.5  & 15.9  & 30.2 & 110 & $8-11$ & 5.8 \\ \hline
C1      & He-N$_2$ & 8.4  & 29.7  & 56.5 & 90  & $2.5-3.6$ & 3.0  \\ \hline
C3      & He       & 17.4 & 61.5  & 117  & 100 & $9$ & 5.6  \\ \hline
\end{tabular}
\label{t:cer}
\end{center}
\end{table}

\section{The old versus new \cer algorithm}

While the \cer hardware used in FOCUS was essentially the same as
E687, a completely new \cer algorithm was written for FOCUS.
This new algorithm will be referred to throughout this
article by the acronym CITADL (for \cer Identification of Tracks
by an Algorithm using Digital Likelihood).  Before describing the new
algorithm, we briefly describe the previous algorithm known as
LOGIC. For a more complete description of this
algorithm see reference \cite{nim}.

Unlike CITADL, whose decision is based on the individual firing pattern of
all 300 cells comprising the FOCUS/E687 \cer system, LOGIC 
based its identification on the overall firing status of C1, C2, and C3.
LOGIC rendered a single identification indicating whether or not the track
was consistent with the electron, pion, kaon, and proton hypothesis.\footnote{Muons 
can only be effectively separated from pions over a narrow
momentum range just below each counter's pion threshold. Both E687 and
FOCUS had a separate muon detection system to provide high quality
muon identification. } This decision was based on the track momentum
and the \cer light observed in the three threshold \cer
counters. A counter was declared ``on'' if any of the cells within the
track's \cer cone fired. A counter was declared ``off'' if no cells
within the cone fired and a minimum number of expected
photoelectrons (typically 2.5) was expected under the pion
hypothesis.\footnote{ In order to save time, LOGIC computed the expected number
of photoelectrons for each cell under the pion hypothesis unless the
track was under the pion threshold for the \cer counter. If a
track was below pion threshold, the light yield was computed under
the electron hypothesis. This allowed the \cer system to help in
the identification of electrons.}  Otherwise the firing status for
that counter was declared unknown and its information was removed from
the final decision. The observed on or off firing status was then
compared to whether or not the counter should have fired under a given
hypothesis.  This prediction was based solely on whether or not the
track momentum exceeded an ``effective'' momentum threshold for that
hypothesis.\footnote{The ``effective'' threshold was slightly higher ($\approx 10\%$) 
than the actual threshold in order to crudely take into account the gradual
rise in the expected light yield with momentum above threshold.}

Although the LOGIC algorithm was very effective at helping to isolate
charm particles in E687, it did have shortcomings.  LOGIC tended
to discriminate against pions when one required positive kaon
and proton identification.  Much of LOGIC's tendency towards light
particle identification was intended given the goal of strongly
suppressing pion backgrounds to the kaons found in Cabibbo
favored charm final states. For example, any cell firing within the
\cer cone sufficed to declare a counter on. But if no cells fired,
a significant amount of predicted light was required before that counter would
be declared off.

An unintended bias was due to accidental firings of the \cer
cells due to ``noise.'' The noise was due to RF noise on cables, tube
noise, and light from untracked, charged particles such as
electromagnetic spray and photon conversions produced in the very
intense photon beam. The electromagnetic noise source could be very
serious for \cer cells located in the center of the system where
occupancies sometimes approached 25--50\%. Both effects tended to
assign \cer light to tracks making them inconsistent with ``heavy''
particles such as kaons and protons.

LOGIC's tendency towards light particle identification both reduced the
efficiency for kaon identification in Cabibbo favored decays and
increased backgrounds for rarer Cabibbo suppressed
decays such as $D^0 \rightarrow \pi^+ \pi^-$ or $D^0 \rightarrow \pi^-
\mu^+ \nu$. In order to suppress the copious backgrounds from $D^0
\rightarrow K^- \pi^-$ or $D^0 \rightarrow K^- \mu^+ \nu$, one would
typically require that the pion had a \cer response which was
inconsistent with that for a kaon. 

While studying Cabibbo suppressed states, we
used $D^0$'s skimmed from a sample of $D^{*+} \rightarrow D^0 \pi^+$
with no \cer cuts to measure the fraction of kaons which passed
the pion cuts. Typically 5\% of kaons were misidentified as pions in
E687 by the LOGIC \cer algorithm. Because of the
inflexibility of the LOGIC algorithm, one would need to redesign the
internal cuts to minimize the misidentification of kaons as pions and
re-run the algorithm from tapes that had the required \cer ADC
information.  Although, in principle, LOGIC could be re-run with other
internal cuts, that was not a practical option since data summary tapes
typically did not contain the \cer ADC information.

CITADL is primarily motivated by the desire to produce a more
flexible \cer identification algorithm than LOGIC.  In fact, the
overall performance of CITADL was significantly better than that of LOGIC,
primarily because CITADL allows for the possibility of accidental
firing of \cer cells. Rather than making a hard decision, on
whether or not a track was consistent with a given hypothesis, CITADL
returned relative likelihoods that the track had a \cer pattern
similar to that expected for the electron, pion, kaon, or proton
hypothesis. One could then, for example, put a minimum cut on the
likelihood ratio that the kaon hypothesis is favored over the pion
hypothesis in order get sufficiently clean kaons to do the required
physics. Unlike LOGIC, very few cuts were
required to be ``hardwired'' in the CITADL algorithm.

Like LOGIC, CITADL only uses the on/off status of \cer cells
rather than their pulse height in identifying particles.  This
decision made the computation of likelihoods simple since a cell's
firing probability is given by the Poisson probability ($1 -
\exp(-\mu)$) where $\mu$ is the expected number of photoelectrons
under the given particle hypothesis.\footnote{This assumes that the
gains and thresholds are set such that a single photoelectron will
produce an ADC count in excess of the threshold required to call a cell on.
Under this assumption a cell will fire unless 0 photoelectrons are
observed when $\mu$ are expected.  The Poisson probability of getting
zero photoelectrons is $\exp(-\mu)$.}

CITADL constructed a log likelihood variable based on the firing
probability for all \cer cells that a given track could
potentially affect all cells within the track's $\beta =1$
\cer cone. Assuming for the moment, that a cell only fired in response to
\cer light, if the cell fired, and $\mu$ photoelectrons were expected,
the log likelihood was incremented by $\log(1 - \exp(-\mu))$; while if
the cell failed to fire the log likelihood was incremented by
$\log(\exp(-\mu))$. Cells which were inside more than
one track's \cer cone were considered ``confused'' and excluded
from the sum. The likelihood returned by CITADL is similar in spirit to the 
traditional continuous likelihood used in fitting. The only difference
is that each event has only two outcomes---on or off. For this reason,
we call it a ``digital'' likelihood.

CITADL returns its identification in the form of $\chi^2$ like
variables which we will call $W_e$ , $W_\pi$, $W_K$, and $W_p$.  They
are defined by $W_i = -2 \sum_j^{\rm cells} \log{P_j}$ where $P_j$ is
the probability for the observed outcome (on or off) for the {\it
j'th} cell under each of the 4 particle hypotheses.  One would
typically require that potential charm decay kaons pass a minimum cut
on a likelihood difference variable such as $\Delta W_K \equiv W_\pi -
W_K$. A large $\Delta W_K$ implies that the kaon hypothesis is
significantly favored over the pion hypothesis. Unlike the case in
LOGIC, there is no need to introduce ``effective'' thresholds , since
the $\mu$ dependence on momentum is explicitly taken into
account. There is also no need to declare a minimum number of
photoelectrons required for a \cer decision. If a very small number of
photoelectrons discriminated the two hypotheses, CITADL returns
likelihood differences close to zero.

\begin{figure}[h!]
\begin{center}
	\includegraphics[width=4.in]{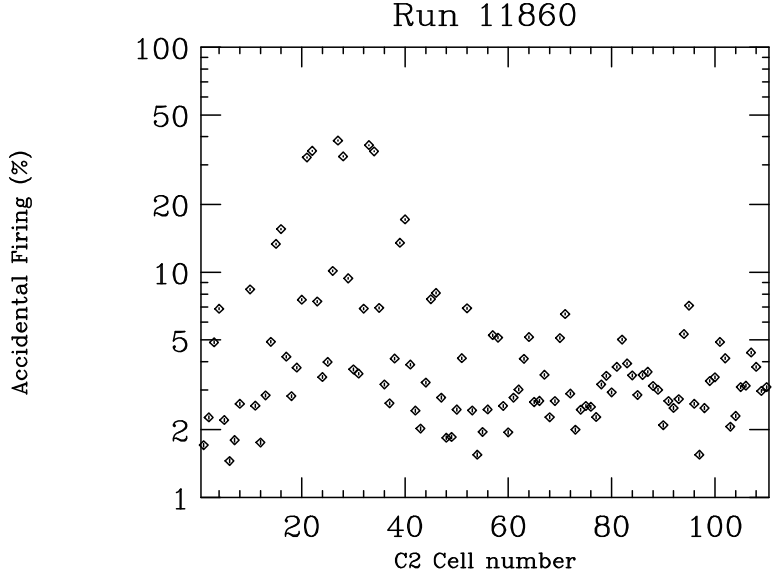}
\end{center} 
	\caption{ The fraction
	of times (in percent) that a cell in C2 fires when no detected
	track's $\beta =1 $ \cer cone impinges on the cell.  These data
	were accumulated over a single run. Although most of the cells
	have an accidental rate of a few percent, cells located near
	the beam axis have accidental rates as high as 40\%. }
\label{accid}
\end{figure}

In computing the log likelihood, we take into account the probability
that a given \cer cell fires accidently due to noise as well as firing
due to a given track. We determined the accidental firing rate by
measuring the fraction of times a \cer cell would fire, even if it
were outside of the $\beta = 1 $ \cer cone of all observed tracks. A
typical plot of the accidental rate as a function of cell number for
one of the runs is shown in Figure \ref{accid}. The accidental rate
varied considerably and for central cells was very large.  It is very
easy to incorporate accidental firing rates in the firing probability.
The prescription is $P_{\mathrm{fire}} = a + (1 - \exp(-\mu)) - a~(1 -
\exp(-\mu))$ where $a$ and $\mu$ are the accidental rate and the
number of photoelectrons expected for the given cell.
We found that $a$ was often proportional to the beam
intensity---especially for cells near the beam axis. CITADL included
this effect as well. The inclusion of realistic
accidental rates significantly improved the performance of our new
algorithm relative to LOGIC.

\section{CITADL Performance}
The very high statistics FOCUS data set provided numerous checks of
the performance of the \cer system and the CITADL algorithm -- often
on a run-by-run basis. These checks used decays into final states of
known daughters. The decays $K_S \rightarrow \pi^+
\pi^-$ provided a very pure and highly copious source of pions,
consisting of 15,000 decays in each of our nearly 6000 runs. This
sample was large enough to provide an accurate photoelectron
re-calibration for nearly all of the 300 cells in the \cer system.

\begin{figure}[h!]
\begin{center}
	\includegraphics[width=4.in]{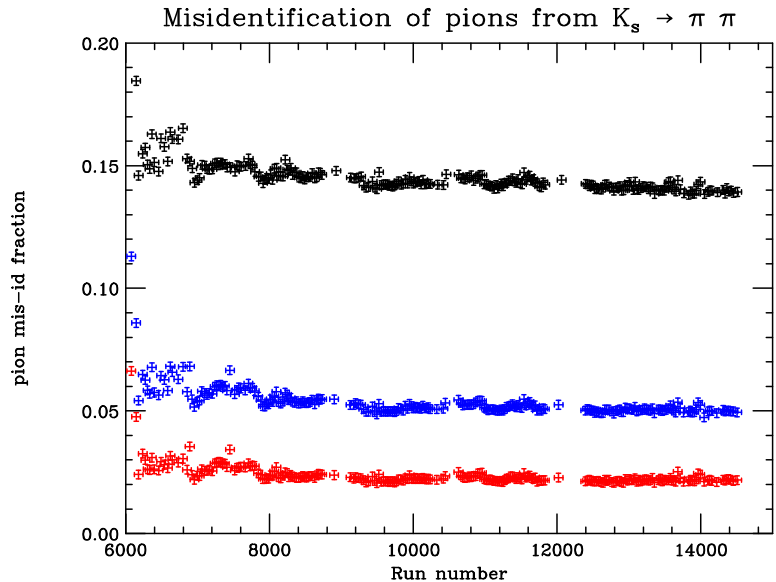} 
\end{center}
\caption{ The
	fraction of times that a pion from $K_S \rightarrow \pi^+
	\pi^-$ is misidentified as a kaon, proton, or electron for three
	different CITADL cuts. Each point is averaged over 25 runs.}
\label{ks_misid}
\end{figure}

\begin{figure}[h!]
\begin{center}
	\includegraphics[width=4.in]{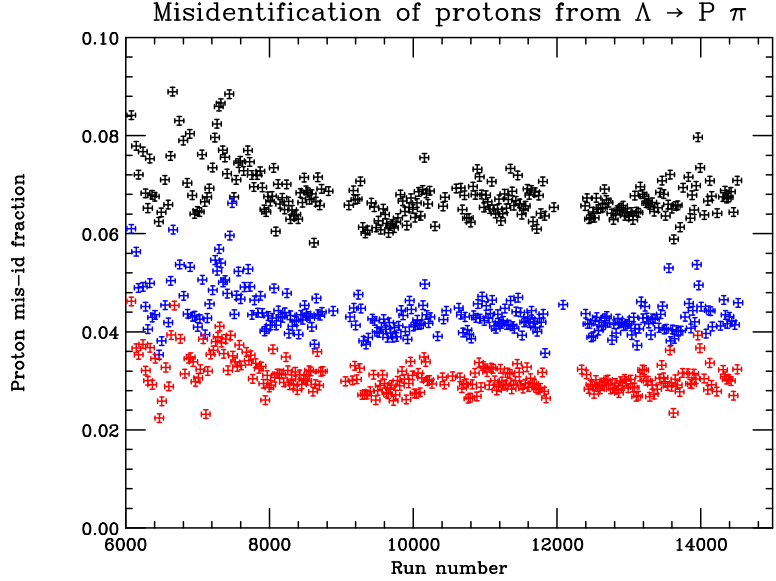} 
\end{center}	
\caption{ The
	fraction of times that a proton from $\Lambda
	\rightarrow p \pi^-$ is misidentified as a light particle for
	three different CITADL cuts. Each point is averaged over 25 runs.}
\label{lam_misid}
\end{figure}

Although not nearly as copious as our $K_S$ sample, the decay $\Lambda
\rightarrow p \pi^-$ provided a clean sample of proton and low
momentum pion decays. \footnote {Reference \cite{cumalat} describes
the method used to reconstruct the $K_S$ and $\Lambda$ topologies in
FOCUS.}  Finally the decay $\phi \rightarrow K^+ K^-$ was used to
measure the \cer identification of kaons on a run-by-run
basis.\footnote{To obtain a clean enough $\phi$ sample to make a
meaningful background subtraction, we required that one of the two
kaons was \cer identified.} The run-by-run fraction of misidentified
daughters from the $K_S$,$\Lambda$, and $\phi$ decays for several \cer
cuts was used as a stability monitor of the \cer system.  Figures
\ref{ks_misid} and \ref{lam_misid} show examples of these
``misidentification'' monitors.

\begin{figure}[h!]
\begin{center}
	\includegraphics[width=4.in]{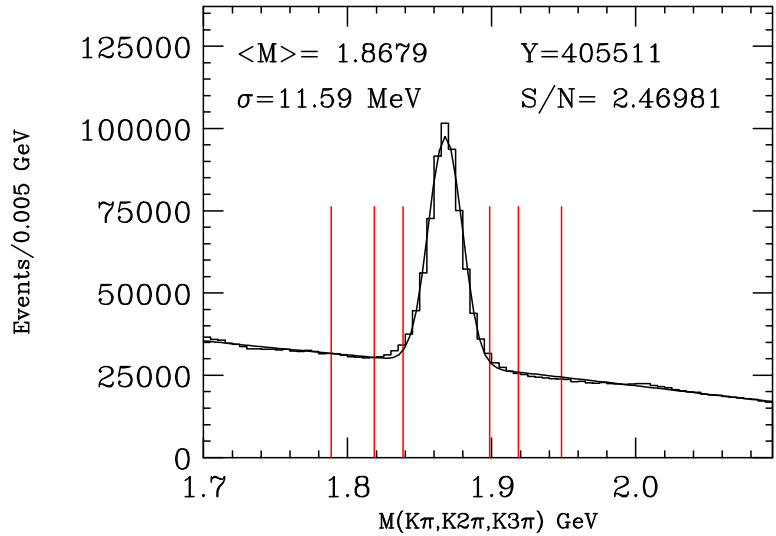} 
\end{center}
\caption{ Invariant mass
  plot for the three golden mode decays $D^0 \rightarrow K^- \pi^+ , K^-
  \pi^+ \pi^+ \pi^-$, and $D^+ \rightarrow K^- \pi^+ \pi^+$. The
  reconstructed $D^+$ mass was shifted by 5~\mevcc so that its peak will
  reconstruct in the same place as the peak of the $D^0$. This data has
  vertex quality and kinematic cuts only. No
  \cer cuts were used.  The vertical lines denote signal and sideband
  regions which will be used to make a background
  subtraction.
  }
\label{side}
\end{figure}
\begin{figure}[h!]
\begin{center}
	\includegraphics[width=5.in]{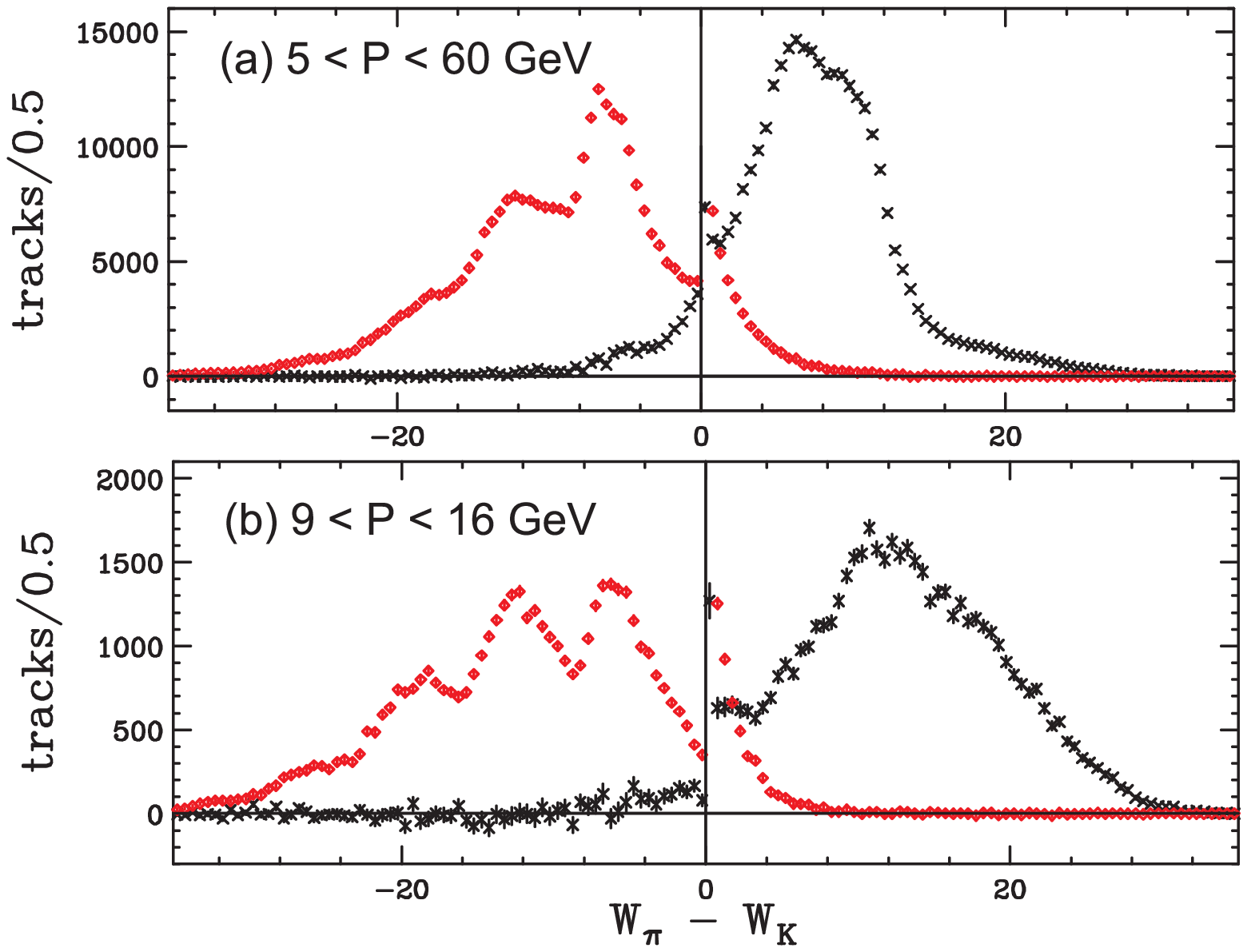} 
\end{center}
\caption{ The log likelihood difference $W_\pi - W_K$ distribution
  obtained from background subtracted kaons (x's) and pions (diamonds)
  from the golden mode charm signal shown in Figure \ref{side}. The
  pion distributions were rescaled to have the same area as the kaon
  distributions. Fig. (a) Tracks with momenta in the range 5
  $<$ P $<$ 60~\gevc. Fig. (b) Tracks with momenta in the range
  9 $<$ P $<$ 16~\gevc. There are off scale spikes in the 0-bin
  consisting of 20,000 and 4,500 events for Fig. (a) and (b).  }
\label{wk}
\end{figure}

We also found that it was possible to use golden mode charm as a
monitor of \cer performance. Figure \ref{side} shows a 405,000 event
golden mode charm sample obtained (using about 75\% of our data)
without any \cer cuts. A selection of cuts on vertex detachment,
isolation, the $D^{*+} - D^0$ mass difference, and momentum were used
to obtain this reasonably clean sample. Also shown are sideband
regions used for background subtraction. Figure \ref{wk} shows the
likelihood difference $\Delta W_K=W_\pi - W_K$ for the kaon and pion
daughters from these background subtracted charm decays for tracks
with two ranges of momentum.  For convenience, we will call the
variable $\Delta W_K \equiv W_\pi - W_K$ ``kaonicity''. A positive
kaonicity implies that a given track is more likely to be a kaon as
opposed to a pion.

Figure \ref{wk}(a) shows the kaonicity distribution for charm
kaons and pions in a momentum range above the pion threshold of C2
(the lowest threshold counter) but below the kaon threshold of C3 
(the highest threshold counter). Outside of this momentum range, the
FOCUS \cer system is incapable of much K-$\pi$ separation and the
kaonicity distribution is strongly peaked near zero.\footnote{CITADL 
offers some slight K-$\pi$ discrimination outside of this range
since it can exploit the momentum dependence of photoelectron yield
beyond the C3 kaon threshold: \emph{i.e.} the threshold is not infinitely
sharp.} Figure \ref{wk}(b) shows the kaonicity distribution in the 
more restricted range from 9 to 16~\gevc. In this range kaon-pion 
discrimination is particularly effective since it lies
above the pion threshold for C1 but below the kaon threshold of C2.

Figure \ref{wk} shows that, even though the likelihoods are constructed
from the discrete firings of \cer cells, the kaonicity
distribution for kaons is reasonably continuous  except
near $\Delta W_K = 0$.  As Figure \ref{wk}(a) shows, averaged over
the accepted charm momentum spectrum, pion backgrounds to kaons can be
very effectively eliminated while still maintaining high efficiency
for charm kaons. A cut just above kaonicity of zero rejects a large
fraction of pions. The fraction of background pions dies
away exponentially with the kaonicity cut above zero.
Over the more restricted range from 9 to
16~\gevc, where cells from both C1 and C2 discriminate pions from kaons,
the $\Delta W_K$ distribution shows a sigificantly larger average kaonicity.
One can make a very stringent kaonicity cut to suppress
pion backgrounds and still maintain good efficiency for real kaons.

The situation for pion identification is essentially the mirror image
of that for kaons. The contamination of kaons into the $\Delta W_K <0$
region falls off exponentially in $\Delta W_K$, while the pion
spectrum extends below $\Delta W_K < -20$. In the region from 9 to
16~\gevc, where both C1 and C2 discriminate pions from kaons, the
average kaonicity of pions becomes significantly more negative
permitting one to make more stringent cuts to reduce
misidentification.

\subsection{Understanding the kaonicity distributions}

We begin by discussing the 
kaonicity distribution for kaons in Figure
\ref{wk}. As described below, we use a 
simplified\footnote{One simplification is that the model assumes that all
pion-kaon discrimination in CITADL is due to cells which have pion
thresholds below the track's momentum but kaon thresholds above the
track's momentum. In fact there is extra discrimination for cells with
kaon thresholds below the track momentum since the probability
that the cell will fire under the kaon hypothesis is less than the
firing probability under the pion hypothesis.} model to conclude 
that the positive
half of the kaonicity distribution is controlled by
twice the total number of
photoelectrons in those \cer cells which discriminate pions from
kaons and the negative half of the spectrum depends on 
the accidental firing rate but is damped exponentially in kaonicity.

When a track is assigned a non-zero kaonicity, there are some \cer
cells that would be expected to fire if the track were a pion and
would be expected to fail to fire if the track were a kaon.  If
$\Delta W_K > 0$, these ``discriminating'' cells did not fire. Let
$\mu$ be the sum of the photoelectrons expected for pions for these
``discriminating'' cells.  Assuming a small accidental rate, the
probability that these cells would fail to fire under the kaon
hypothesis is essentially 1, and under the pion hypothesis is
$\exp(-\mu)$. The kaonicity is defined in terms of the probability
that the light pattern agrees with the kaon hypothesis divided by the
probability that the pattern agrees with the pion hypothesis or $\Delta
W_K = -2 \log(P_\pi/P_K)$. After inserting these non-firing
probabilities, we have $\Delta W_K = 2 \mu$.  According to this model,
the maximum kaonicity of $\approx 30$ means that the FOCUS \cer system
provides at most 15 photoelectrons which discriminate between kaons
and pions. In the momentum region between 9 and 16~\gevc, Figure
\ref{wk}(b) shows that the yield of discriminating photoelectrons is
much larger than over the full range from 5~\gevc to 60~\gevc. This is
because pions with momenta in this range should fire the cells of both
C1 and C2 whereas kaons should not. The photoelectron yield for
$\beta=1$ tracks in C2 is typically in excess of 11---larger than that
for C1 or C3.

If CITADL assigns a track $\Delta W_K <0$, there must be
discriminating cells which fired making the pion hypothesis more
likely than the kaon hypothesis. For real kaons, such as those
displayed in Figure \ref{wk}, this can only happen due to accidental
firing. Denote the probability that noise fires a discriminating cell
by $a$.  In the limit where there is a reasonable number of
discriminating photoelectrons, the probability that the cells will
fire under the pion hypothesis will approach 1. Hence if a cell
accidently fires for a kaon, CITADL will report a kaonicity of $\Delta
W_K = -2 \log(P_\pi/P_K) = 2 \log(a)$ where $a$ is the accidental
firing rate. The probability that a kaon track will actually fire a
discriminating cell will of course be $a = \exp(\Delta W_K/2)$.  One
therefore expects the roughly exponential fall off in negative
kaonicity for real kaons which is observed in Figure
\ref{wk}.\footnote{ The distribution of negative kaonicities for kaons
also depends on the distribution of noise rates for cells in the
\cer system. Deviations from an $\exp(\Delta W_K/2)$ distribution
are therefore expected since the distribution of
accidentals is nonuniform as shown in Figure \ref{accid}.}  To
summarize, a positive kaonicity value is essentially twice the number
of photoelectrons which discriminate kaons from pions at the momentum
of the kaon; while the negative half depends on the distribution of
accidental firing rates and is suppressed by a factor of $\exp(\Delta
W_K/2)$.

We next turn to a discussion of the $\Delta W_K$ distribution obtained
for the pions shown in Figure \ref{wk}. For negative kaonicities
($\Delta W_K<0$), the kaonicity distribution exhibits considerable
structure but when $\Delta W_K >0$ it dies exponentially with
kaonicity. The exponential fall-off in the $\Delta W_K >0$ region is
due to discriminating cells not firing for the pion thus causing
CITADL to prefer the kaon hypothesis.  For pions, this should happen
with a probability of $P_\pi = \exp(-\mu)$ while for kaons it will
occur with $P_K=1$. Such tracks will therefore be assigned a kaonicity of
$\Delta W_K = -2 \log(P_\pi/P_K) = 2 \mu$. We thus expect a kaonicity
distribution for $\Delta W_K >0$ given by the product of the spectrum
of discriminating photoelectrons times the probability of the pion not
firing the discriminating cells. This leads to a nearly exponential
distribution since the probability of a pion not firing the
discriminating cells is given by: $\exp(-\mu) = \exp(-\Delta W_K/2)$.
In fact, the kaonicity spectrum of pions in the region 
$0.5 < \Delta W_K < 10$ is well fit
to the form $\exp(- 0.4~\Delta W_K)$. 

We next consider the $\Delta W_K < 0$ half of the kaonicity spectrum
for pions. In this region, the pion hypothesis is favored over the
kaon hypothesis if the discriminating cells fired.  Assuming a
relatively large number of discriminating photoelectrons, the
probability the cells will fire is close to $P_\pi = 1$. Under the
kaon hypothesis the cells would only fire due to accidentals which
would occur with a probability of $P_K = a$. The pion would then be
assigned a kaonicity of $\Delta W_K = -2 \log(P_\pi/P_K) = 2 \log
(a)$.  If several discriminating cells fire, the kaonicity
distribution will be incremented by several multiples of $2
\log(a)$. Indeed several peaks are present in the $\Delta W_K < 0 $
spectrum of Figure \ref{wk}(b) which appear at multiples of
approximately 6.25 which implies a typical accidental rate of $a =
\exp(-6.25/2) = 0.044$. This estimated $a$ is consistent with the
typical accidental rate shown in Figure \ref{accid}.

To summarize, the $\Delta W_K > 0$ spectrum for pions is controlled by
the number of discriminating photoelectrons and is damped by a 
Poisson inefficiency factor of $\exp(-\Delta W/2)$ while the
$\Delta W_K < 0$ region is controlled by multiples of 
twice the log of the accidental firing rate. 

This model, among other things, explains why the pion-kaon separation
in the momentum range of Figure \ref{wk} (b) is so much better than
over the complete momentum spectrum. Because C1 and C2 both
discriminate in this narrow momentum range, there are more \cer
cells available to
discriminate between the pion-kaon hypothesis. Hence there are more
discriminating photoelectrons which increases the average kaonicity
for kaons, and more multiples of $-2 \log(a)$ which
further decreases the (negative) kaonicity distribution for pions.

\subsection{Using \cer Information to Reduce Charm Backgrounds}

\begin{figure}[h!]
\begin{center}
	\includegraphics[width=4.in]{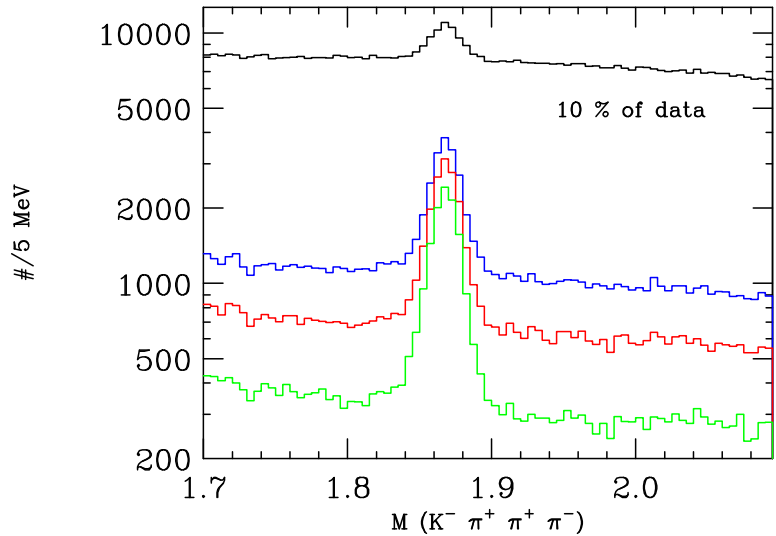} 
\end{center}
\caption{
Illustration of the effectiveness of \cer cuts in reducing
backgrounds to $D^0 \rightarrow K^- \pi^+ \pi^+ \pi^-$. 
Note the logarithmic scale. The upper
curve has no \cer cuts. The second histogram requires $\Delta W_K > 0$.
The third requires $\Delta W_K > 2$. The fourth histogram
requires $\Delta W_K > 2$ and $\Delta W_\pi > -2$. A considerable 
improvement in the signal to noise is evident with only moderate loss
in efficiency. The fitted signal yields in these plots are 15307, 12783, 11699, and 9996 
respectively.}
\label{k3pi}
\end{figure}

As Figure \ref{side} shows, it was indeed possible to get reasonably
clean charm signals without the use of \cer information. However, many
FOCUS analyses employed \cer cuts as an effective way of increasing
signal to noise, while maintaining reasonable efficiency. Figure
\ref{k3pi} illustrates the effectiveness of kaon and pion \cer cuts
for $D^0 \rightarrow K^- \pi^+ \pi^+ \pi^+$ events selected by
requiring that the secondary to primary vertex detachment exceeded 9
standard deviations. No \cer cuts were used in the initial
selection. The kaon cut is on ``kaonicity'' or the log likelihood
difference $\Delta W_K \equiv W_\pi - W_K$ discussed previously. The
pion cut is based on a pion consistency variable which we will call
``piconicity'' or $\Delta W_\pi \equiv W_{\rm min} - W_\pi$ , 
where $W_{\rm min}$ is smallest negative log likelihood
of the 4 particle hypotheses.  The
$\Delta W_\pi$ cut is placed on all $D$ decay pions and is meant to
ensure that no pion being considered as a charm daughter is grossly
inconsistent with the pion hypothesis.\footnote{We generally use a
consistency cut rather than demanding that the pion is favored over
both the kaon and electron hypothesis since the momentum range at
which pions can can be distinguished from electrons is below 17~\gevc
for tracks traversing all three \cer counters and below 8.5~\gevc for
3 chamber pions which traverse only C1 and C2.}  A cut such as $\Delta
W_\pi > -2$ means that none of the other 3 particle hypotheses is
favored over the pion hypothesis by more than a factor of $\exp(2/2) =
2.71$.  For the $D^0 \rightarrow K^- \pi^+ \pi^+ \pi^+$ sample the
requirement that $\Delta W_K > 0$ preserves 84\% of the yield while
increasing the signal to noise by a factor of 6.2. The more stringent
$\Delta W_K > 2$ and $\Delta W_\pi > -2$ preserves 75\% of the uncut
signal yield but increases the signal to noise by a factor of 16.

\begin{figure}[h!]
\begin{center}
	\includegraphics[width=4.in]{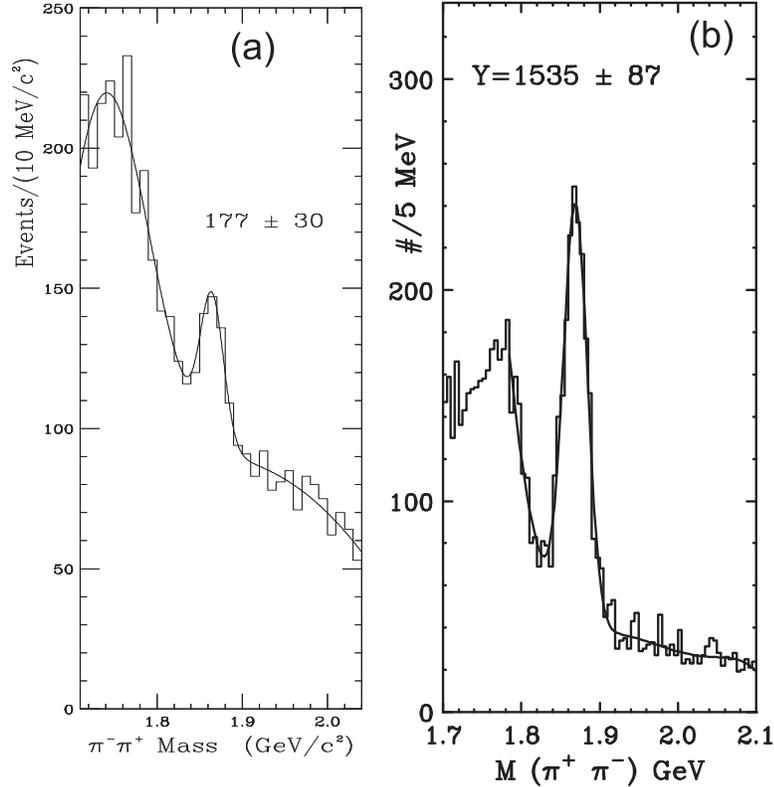}
\end{center}
\caption{
(Left) 
E687 signal for $D^0 \rightarrow \pi^+ \pi^-$ from Reference \cite{e687pipi}.
The massive distortion in the background at lower masses is due to
contamination from misidentified $D^0 \rightarrow K^+ \pi^-$. 
(Right) FOCUS signal for $D^0 \rightarrow \pi^+ \pi^-$ from half of our
data set.
}
\label{pipi}
\end{figure}

One of the goals of the CITADL algorithm was to be much more efficient
than LOGIC in suppressing the number of kaons which are misidentified
as pions to enable us to more effectively study Cabibbo suppressed
decays. An example of such a process is the $D^0 \rightarrow \pi^+
\pi^-$ which is plagued by a large background from
misidentified $D^0 \rightarrow K^+ \pi^-$ decays which occur with
a branching ratio that is approximately 
25 times larger than that of $D^0 \rightarrow \pi^+
\pi^-$.  Figure \ref{pipi} compares the dipion mass spectrum from the
published E687 signal to a version from half of the FOCUS data set.
The E687 sample used LOGIC. The FOCUS sample required $W_K - W_\pi >
3$ for both pions in order to significantly reduce $D^0 \rightarrow
K^+ \pi^-$ contamination. Much of the improvement in event yield is
due to the fact that FOCUS took roughly a factor of 15 times the E687
data set. We have also required that the dipion vertex be outside of
the FOCUS target material to further increase our signal to noise
relative to E687. However, the CITADL algorithm is responsible for the
significant reduction in FOCUS data of the reflection from
misidentified $D^0 \rightarrow K^+ \pi^-$ relative to the $D^0
\rightarrow \pi^+ \pi^-$ signal compared to what was achievable in
E687.

\begin{figure}[h!]
\begin{center}
	\includegraphics[width=4.in]{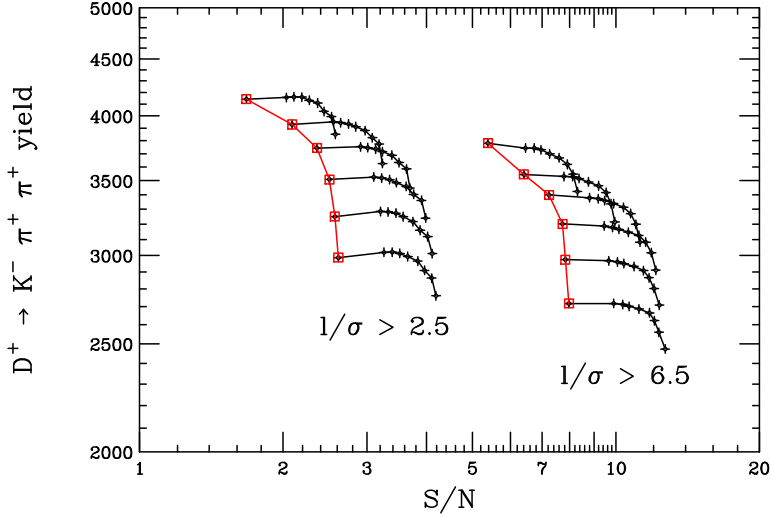} 
\end{center}
\caption{ Illustration of the effectiveness of \cer cuts in reducing
backgrounds to $D^+ \rightarrow K^- \pi^+ \pi^+$ which verticize
outside of the target microscrips and target material.  We form a
``cut tree'' by plotting the signal yield versus S/N for two different
detachment cuts and several cuts on kaonicity and pion consistency
(piconicity).  The kaonicity cuts (the main trunks) range from $\Delta
W_K > 0.5,1,2,3,4,5$. The piconicity cuts (the branches) are ``no cut'',
$\Delta W_\pi > -10, -9, ..., -3$. Only 2 \% of the our complete data
was used for this plot.  }
\label{tree}
\end{figure}

Figure \ref{tree} is an example of a plot used to gauge
the effectiveness of a set of \cer cuts on the pions and kaons
from a very small sample of $D^+ \rightarrow K^- \pi^+ \pi^+$
decays. The data satisfied our standard skim cuts for this mode: a
good quality vertex intersection (CL $>$ 1 \%) , a kaonicity cut of $\Delta
W_K > 0.5$,  and a secondary to primary detachment exceeding 2.5
standard deviations ($\ell/\sigma > 2.5$).  This particular plot used
the sample of $D^+$ decays which verticized outside of the
target material and target microstrip system to remove
backgrounds from multiple interactions.  We show the yield versus
signal to noise for 2 detachment cuts, and a sequence of \cer cuts
on the kaons and pions.\footnote{Both the yields and signal to noise
were based on fits to a Gaussian signal over a polynomial
background. We define the signal to noise ratio as the ratio of the
fitted number of signal events at the peak over the fitted number of
background events at the peak mass.} 
Figure \ref{tree} shows that the ``piconicity cut'' is essentially as
effective a cut as the kaonicity cut. Figure \ref{tree} also shows
that the \cer cuts increase the signal to noise by a nearly 
constant factor at the two detachment cuts being considered.

\section{Calibration and Monitoring}
\begin{figure}[h!]
\begin{center}
	\includegraphics[width=4.5in]{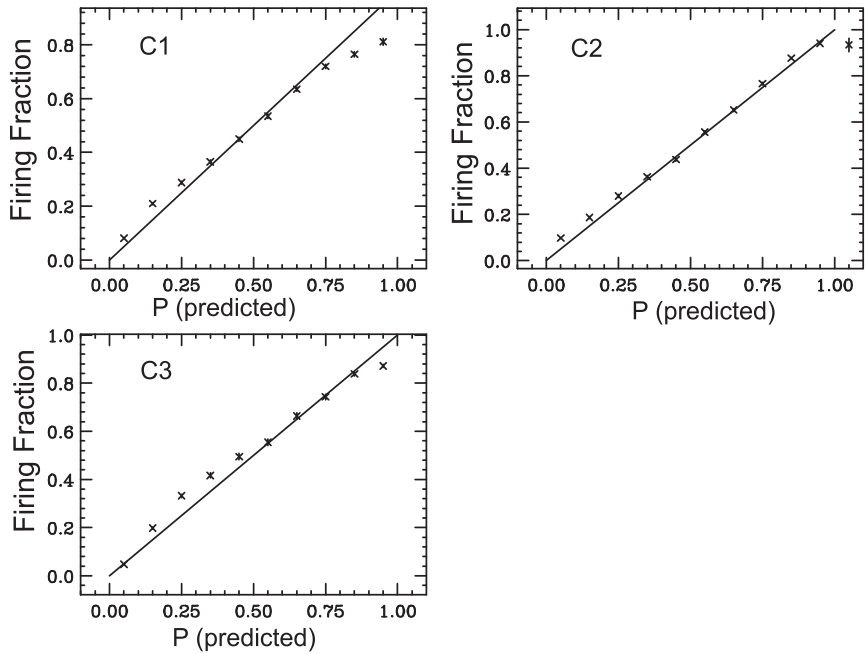} 
\end{center}
\caption{ Plots of
the actual firing rate as a function of the predicted firing rate for
all cells within the given counter. We use pions from the decay $K_S
\rightarrow \pi^+ \pi^-$.  This set of plots was obtained in run
13955.}
\label{weather}
\end{figure}

We made a large number of plots while reconstructing our data to
monitor \cer system performance. Examples of such plots which we have
already discussed include Figures \ref{accid}, \ref{ks_misid}, and
\ref{lam_misid} which serve as monitors of accidental rates, and Vee
($K_S$ or $\Lambda$) daughter misidentification.  We found that our
most powerful calibration tool used pions from $K_S \rightarrow \pi^+
\pi^-$ since roughly 15,000 clean $K_S$'s were reconstructed during
each of our $\approx 6000$ data runs.  These pions were used to study
how well we could predict the firing rate for cells. An example of
such a study is shown in Figure \ref{weather} which plots the observed
average firing rate versus the predicted.  These particular plots are
summed over all cells in each \cer counter. They represent the
statistics on a single run. The prediction depends on the assumed cell
geometry, the accidental firing rates, the relative quantum efficiency
of each tube, and the validity of the analytic model used to quickly
compute the fraction of \cer light falling within the cell
boundary. This model was known to have problems for cells on the
planar mirror boundaries of C1 and the planar mirror apex of C2. It is
clear from Figure \ref{weather} that the light predictions were
imperfect.  We believe that the impact of these imperfections was
slight on overall \cer identification.

We worked diligently to insure that CITADL used good information on the
performance of each of the 300 cells comprising the FOCUS \cer
information. The two critical ingredients are the photoelectron yield
for a $\beta = 1$ track and the ``accidental'' firing rate.
We found that the
performance of some photomultipliers varied significantly
during the 12 month run. Occasionally this was due to the
maintenance of the \cer system, such as changing bases and phototubes. 
More often, small
shifts in the ADC pedestals would create a dramatic apparent increase
in the ``noise'' level of the tube, which could be easily corrected by raising
the minimum ADC count required by CITADL to call a cell ``on.''  Although
special calibration runs were taken in order to understand the \cer system,
we found that the best monitoring of the \cer system was obtained
through regular data taking. In this section, we describe some of these
\emph{in situ} calibration and monitoring methods.

\subsection{Photoelectron Calibration and Monitoring}

\begin{figure}[h!]
\begin{center}
	\includegraphics[width=4.in]{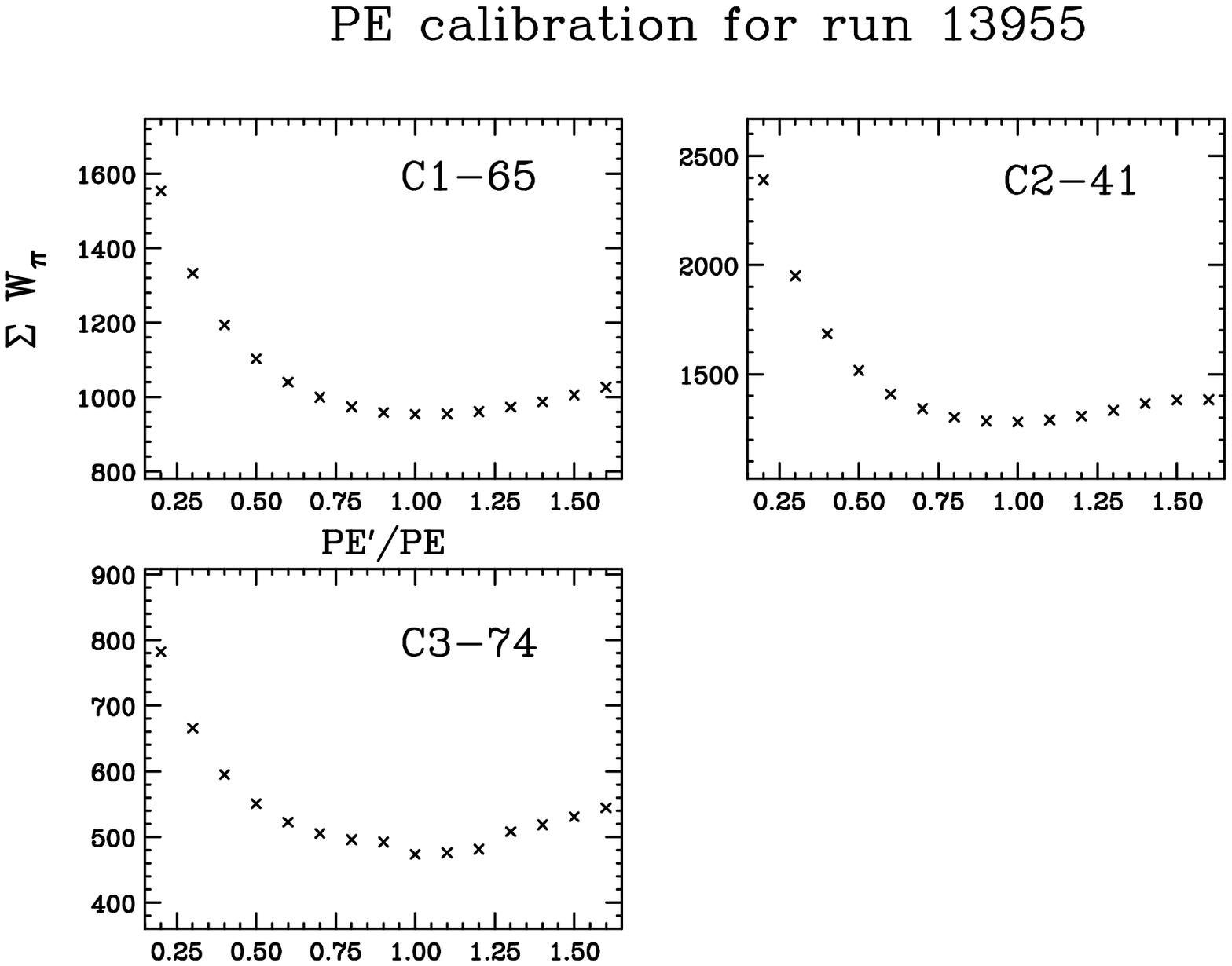} 
\end{center}
\caption{ 
Calibration curves for three cells in the \cer system.  We plot
 $\sum W_\pi$ for pion daughters from background subtracted 
$K_S \rightarrow \pi^+ \pi^-$ decays which strike the cell summed over all
such pions in run 13955. This likelihood is computed for 15 assumed 
photoelectron ratios relative to the $\beta =1$ yield used by CITADL.  
The example cells are in the inner section of each \cer counter
}
\label{parabs}
\end{figure}

We developed a powerful way of monitoring the photoelectron yield as a
function of time for nearly all of the 300 cells comprising the FOCUS
\cer system. This calibration method fit for
the $\beta = 1$ photoelectron yield of each cell 
by minimizing CITADL likelihood 
$W_\pi$ for pions from $K_S \rightarrow \pi^+ \pi^-$.
The $W_\pi$ was incremented for each $K_S$ pion
daughter that was predicted to leave at least 0.1 photoelectron in a
given cell. The likelihood was incremented using background subtraction
weight of 1 if the $K_S$ mass was in the signal region and $- 1$ if the
mass fell in symmetrically placed, half width sidebands. Separate $\sum
W_\pi$ sums were computed for different assumed $\beta =1 $ photoelectron
yields for the given cell which ranged from 20\% to 160\% of the
nominal photelectron yield.  Typical log likelihood versus
photelectron ratio plots obtained in a single run are shown in Figure
\ref{parabs}. With the statistics available in a single run one could
get an adequate re-calibration of the inner \cer cells which are
struck most often.

\begin{figure}[h!]
\begin{center}
	\includegraphics[width=6.in]{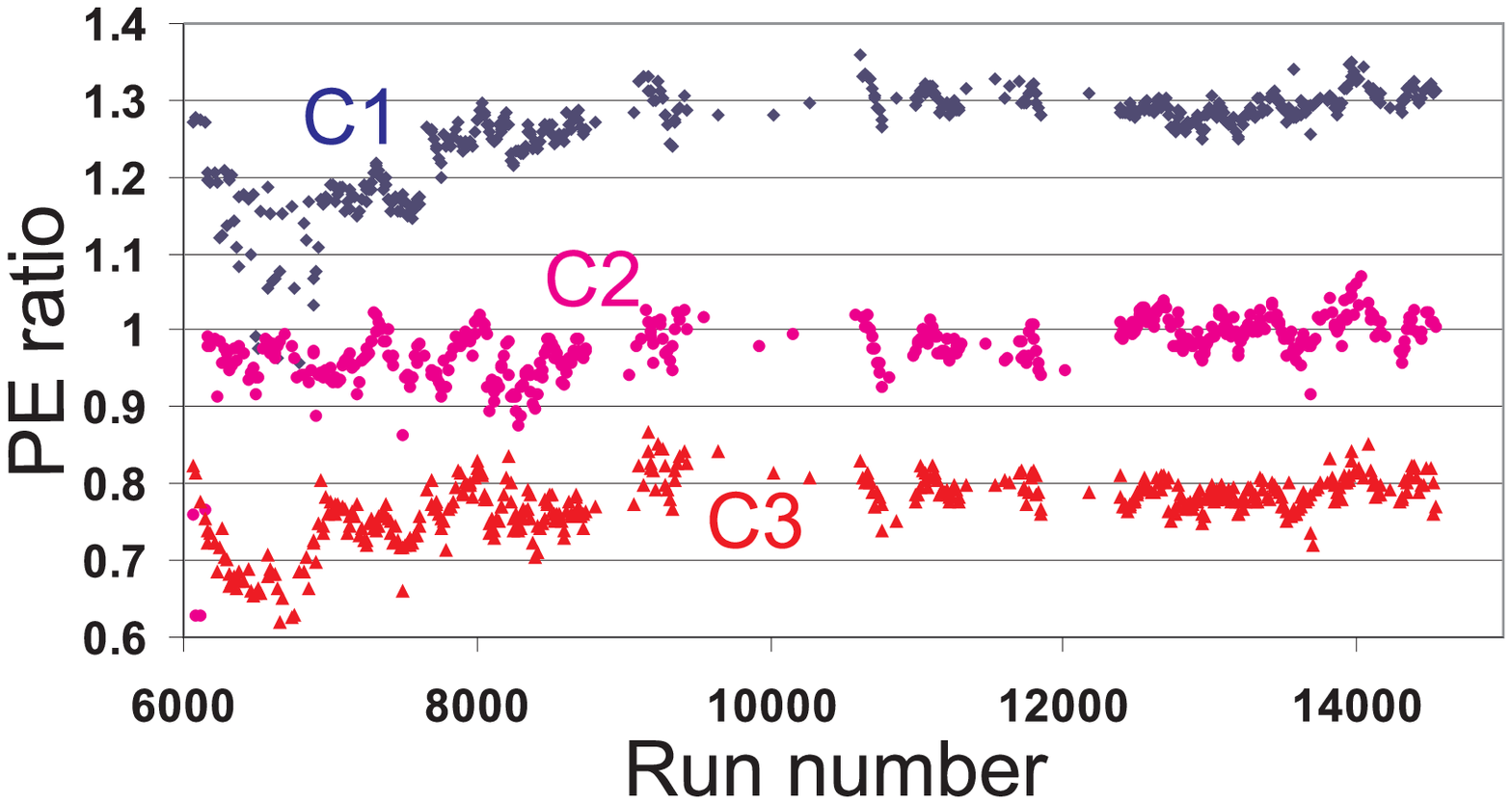}
\end{center}
 \caption{ Plot of the relative photoelectron yield for all three
	counters as a function of run number. Each point is a 10 run
	average and the data for each counter are  offset vertically
	for clarity.  The photelectron yield relative to that assumed
	in the calibration is always within 20\% of unity until run
	9000 and within 10\% thereafter.}
\label{barometer}
\end{figure}

As a stability monitor, we also summed the $W_\pi$ from all of the
$K_S$ pions in a given run over all of the cells in our three \cer
counters. The minima of these grand likelihoods
are plotted as a function of run
number in Figure \ref{barometer} and give an overall scale
factor for each of the three counters . Apart from some photoelectron
fluctuations unique to C1 in the early running, Figure \ref{barometer}
shows that the average photoelectron yield from all three \cer counters
tended to fluctuate together in a way which we learned was correlated
with changes in barometric pressure.

\subsection{Noise Calibration and Monitoring}
\begin{figure}[h!]
\begin{center}
	\includegraphics[width=3.5in]{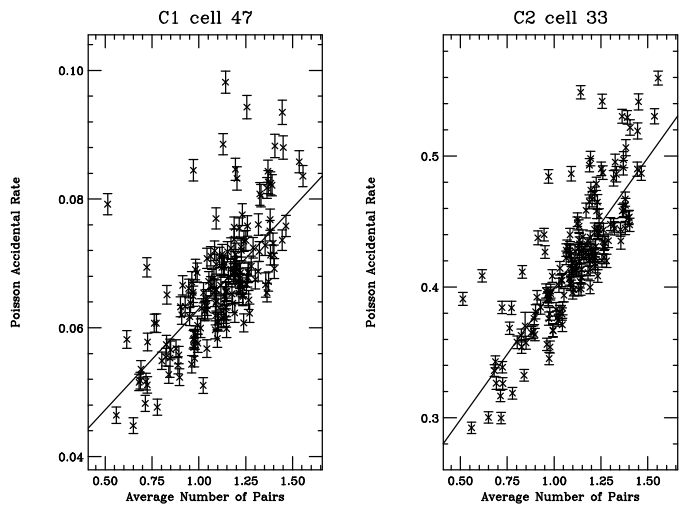}
\end{center}
 \caption{ Plots of
	Poisson accidental rate versus instantaneous beam intensity
	for two \cer cells.  }
\label{intensity}
\end{figure}
CITADL makes direct use of the observed accidental firing rate for
\cer cells which is generally $\approx 2-5 \%$ for most cells but can
be quite high ($\approx 40 \%$) for cells close to the beam axis. We
believe much of this noise for central cells is due to the high rate
of $e^+ e^-$ pairs that accompany our hadronic photoproduced events.
Most of the very ``noisy'' cells showed an accidental rate which was
roughly proportional to the instantaneous beam intensity. Figure
\ref{intensity} illustrates this point for two \cer cells. A few words
are in order. As a measure of the accidental rate, we use a variable
called the ``Poisson Accidental Rate''.  This ``Poisson'' rate
($\mu_a$) is related to the accidental rate $a$ (or fraction of times
a cell fires when no track is pointing at it) via $1 - \exp(-\mu_a)
\equiv a$. We assumed in our treatment that $\mu_a$ rather than $a$
was linear in the beam intensity.\footnote{For small $a$ , $a \approx
\mu_a$ but clearly as $a \rightarrow 1$, $a$ cannot continue to grow
linearly with intensity.} We found that the average number of embedded
pair tracks per triggered event formed a very convenient monitor of
our average instantaneous beam intensity.  Embedded pair tracks are
from a Bethe-Heitler beam conversion which happen to lie within the
resolving time of the chambers and microstrips when triggered by one
of our event triggers. Pair tracks are easy to identify since they are
consistent, within their expected multiple scattering, with being
produced along the beam axis.\footnote{The average number of embedded
pair tracks was $\approx 1$ per event throughout much of the FOCUS
running owing to our very high intensity.}

We found that the average number of 
pair tracks in coincidence with an event trigger was a good measure of 
the instantaneous beam intensity which included the sometimes dramatic effects
of spill non-uniformity. We used linear fits of the Poisson accidental rate
versus $\langle\mathrm{pair~tracks} \rangle$, such as those in Figure \ref{intensity},
to model the noise response for each of the 300 \cer cells. Often several
parameterizations were made for a given cell to cover run dependent changes
in the accidental rate. 

\begin{figure}[h!]
\begin{center}
	\includegraphics[width=5.in]{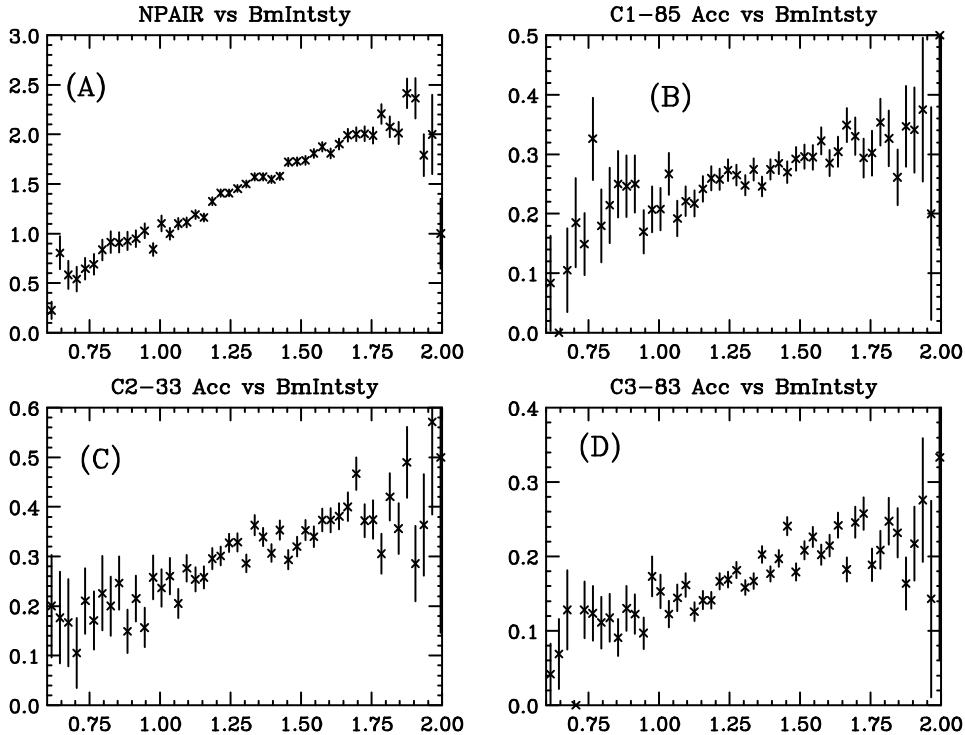} 
\end{center}
\caption{Rate
	dependences of several quantities versus the scaler deduced
	intensity monitor within a single accelerator spill. Figure
	(a) shows the average number of embedded pair tracks versus
	the beam intensity. Figures (b), (c), and (d) show the
	accidental rate in three \cer cells versus the beam
	intensity.  }
\label{scaler}
\end{figure}

In fact the beam intensity varied significantly between the spills and
even within a spill.\footnote{Typically we had one spill of protons
every minute which lasted for about 20 seconds. The first 500
microseconds of this spill was ``fast'' extraction for neutrino
experiments. A typical run lasted 40 minutes.} To optimize \cer
algorithm performance, we devised a method to estimate the beam
intensity directly before each recorded event and use this intensity
measurement to estimate the accidental rate for each cell.  This
method used several scalers which were recorded on our raw event
tapes during each second level, event trigger. One of these scalers
counted the accelerator RF clock, the others counted the hits in two
scintillation counters used in the first level trigger.  The scalers
were effectively reset each time we read an event out.  The time rate
of change of either of these two scintillation counters formed a
direct measurement of the beam intensity right before the actual
event. Figure \ref{scaler} shows that both the average number of
embedded pair tracks, and the accidental \cer cell firing rates are
strongly correlated with the scaler derived beam intensity even within
a single spill of Run 7185.

\section{Summary and Conclusions}

In this article we describe a likelihood based \cer algorithm used to
identify charged particles in the FOCUS charm photoproduction
experiment. FOCUS used three multicell \cer counters, operating
in threshold mode.  We believe that this algorithm may prove useful in
future experiments employing threshold \cer counters in high rate
environments. This article describes the CITADL algorithm,
illustrates its effectiveness on charm signals, and discusses the method
we used for continuous monitoring of each cell's photoelectron yield and
noise.

Although the pulse heights for all of the firing \cer cells in a
given event were recorded, our algorithm did not use this pulse height
information.  The CITADL algorithm was based on the probability that
\cer cells uniquely associated with a given track either fired or
failed to fire.  Because there were only two outcomes per \cer cell,
we say the algorithm returned a ``digital likelihood'' for a given
particle hypothesis. Given the large number of cells with different
accidental rates and photoelectron yields comprising the FOCUS \cer
system,the ``digital'' likelihood provided an essentially continuous
identification variable.

By using only the on/off status of the \cer cells, we found that it
was possible to include the effects of ``accidental'' firing due (in
our case) to the untracked electromagnetic debris in regions close to
our photon beam.  Essentially we combined the probability of a cell
firing due to \cer light, with an accidental firing probability using
De Morgan's Law.  The accidental probability for each \cer cell was
parameterized in terms of an intrinsic accidental rate and a
contribution proportional to the instantaneous beam intensity. By
realistically including the accidental rate in our likelihood, we
substantially improved our ability to identify light particles (such
as pions) over the less sophisticated algorithm used in our previous
charm photoproduction experiment, E687. Interestingly enough, the
ability to positively identify pions with high efficiency and low kaon
contamination, proved useful in significantly increasing our signal to
noise even in Cabibbo favored charm decays.

The goal of the CITADL algorithm was to provide flexible
identification with a broad efficiency versus misidentification curve.
The flexibility of this algorithm has proven very useful in assessing
systematic errors due to misidentified charm reflections in our recent
studies of Cabibbo suppressed and doubly suppressed charm
decays.\cite{kklife}-\cite{link}

We wish to acknowledge the assistance of the staffs of Fermi National
Accelerator Laboratory, the INFN of Italy, and the physics departments
of the collaborating institutions. This research was supported in part
by the U.~S.  National Science Foundation, the U.~S. Department of
Energy, the Italian Istituto Nazionale di Fisica Nucleare and
Ministero dell'Universit\`a e della Ricerca Scientifica e Tecnologica,
the Brazilian Conselho Nacional de Desenvolvimento Cient\'{\i}fico e
Tecnol\'ogico, CONACyT-M\'exico, the Korean Ministry of Education, and
the Korean Science and Engineering Foundation.


\begin{thebibliography}{10}
\bibitem{nim} E687 Collab., P. L. Frabetti {\it et al.,}
Nucl. Instrum. Methods Phys. A 320 (1992) 519.
\bibitem{cumalat} FOCUS Collab., J. M. Link {\it et al.,}
Submitted to Nuclear Instruments and Methods.
\bibitem{e687pipi} E687 Collab., P. L. Frabetti {\it et al.,}
Phys. Lett. B321 (1994) 295
\bibitem{kklife} FOCUS Collab., J. M. Link {\it et al.,}
Phys. Lett. B485 (2000) 62
\bibitem{cp} FOCUS Collab., J. M. Link {\it et al.,}
Phys.Lett.B491 (2000) 232
\bibitem{link} FOCUS Collab., J. M. Link {\it et al.,}
Phys.Rev.Lett. 86 (2001) 295 
\end{thebibliography}
\end{document}